\documentclass[prb,twocolumn,amssymb]{revtex4}
\usepackage{graphicx}

\begin{document}
 \title{Effect of the surface curvature on the magnetic moment and persistent currents
 in two-dimensional quantum rings and dots}
 \author{D. V. Bulaev}
 \email{bulaevdv@mrsu.ru}
 \author{V. A. Geyler}
 \author{V. A. Margulis}

 \affiliation{Mordovian State University, Saransk, 430000 Russia}

\date{\today}

\begin{abstract}
The effect of the surface curvature on the magnetic moment and persistent currents in
two-dimensional (2D) quantum rings and dots is investigated. It is shown that the surface
curvature decreases the spacing between neighboring maxima of de Haas --- van Alphen (dHvA) type
oscillations of the magnetic moment of a ring and decreases the amplitude and period of Aharonov
--- Bohm (AB) type oscillations. In the case of a quantum dot, the surface curvature reduces the
level degeneracy at zero magnetic fields. This leads to a suppression of the magnetic moment at
low magnetic fields. The relation between the persistent current and the magnetic moment is
studied. We show that the surface curvature decreases the amplitude and the period of persistent
current oscillations.
\end{abstract}

\pacs{73.20.At, 73.23.Ra, 75.75.+a}

\maketitle

\section{Introduction}
A metallic ring of mesoscopic dimension in an external magnetic field is known to exhibit a wide
variety of interesting physical phenomena: AB effects \cite{AB,KBH}, quantum Hall effects
\cite{H}, persistent currents \cite{TI99,AHK}, the Berry phase \cite{BK}, and spin-orbit effects.
\cite{MEG}

One of the most important factors that cause complications in real experiments is the finite
width of rings. A.~Lorke and co-workers \cite{LLG} showed that even in very small nanoscopic
quantum rings, occupied with one or two electrons, there are some electron modes, corresponding
to different radiuses of electron orbits in the ring. In a ring with finite width, not only are
multiple channels effects important, but the penetration of the uniform magnetic field into the
conducting region of the ring also plays a significant role. \cite{BL-95,TI99,TI96,AHK,MShT} For
example, the penetration of a magnetic field into the conducting region can result in aperiodic
oscillations of the persistent currents \cite{KBH,TI96,TI99,MShT} and the breakdown of the simple
linear relation between the persistent current and the magnetization. \cite{TI99}

There are many theoretical models which take into account the
finite width of rings \cite{BL-95,TI99,TI96,AHK,MShT}, such as the
model of a 2D ring with a hard-wall confinement potential
\cite{Klama,AHK}, a parabolic potential \cite{ChP,PJA}, and a
finite square-well potential. \cite{TS} Recently, the current-spin
density-functional theory has been employed to study the combined
effects of the confinement, Coulomb interactions, a spin
polarization, and a magnetic field in a quantum dot and a quantum
ring. \cite{PS,LG01} The above mentioned models allow only
numerical analysis. There are some exactly solvable models for 2D
quantum rings. \cite{BL-95,TI96,TI99,MShT} Experiments on the
spectroscopy of nanoscopic rings showed that a parabolic potential
describes lowest electron states in a quantum ring very well.
\cite{LLG} Note that confining potential proposed in
Refs.~\onlinecite{BL-95,TI96} is in very good agreement with the
experiment also.

Quantum interferences lead to a number of new phenomena in the transport properties of
nanostructures. However, the phase coherence conservation of the electron wave functions in the
whole sample can also affect the equilibrium properties of the system. N.~Bayers and C.N.~Yang
\cite{BayersYang} were the first to show that an isolated normal-metal ring threaded by a
magnetic flux carries an equilibrium current at finite temperature as long as the electron phase
coherence is preserved. The work by M.~B\"{u}ttiker, Y.~Imry, and R.~Landauer \cite{BIL},
predicting persistent currents in one-dimensional (1D) disordered loops, renewed the interest in
the topic. This interest is heightened \cite{TI99,VSR} by recent advances in submicrometer
physics \cite{MCB} that have brought the effect into reach of experimental investigation. It was
shown that in very thin quantum rings the persistent current is a periodic function of the
magnetic flux with a period $\Phi_0=hc/|e|$ (the AB effect). The persistent current $I$ in this
nanostructure is simply related to the magnetic moment $M$ of the ring by $I=cM/S$, where $S$ is
the area of the ring. However, for a wide ring, in addition to orbital modes of electron motion
in the ring there are also radial modes. W.-C.~Tan and J.C.~Inkson \cite{TI99} showed that the
presence of the radial modes leads to aperiodicity and complication of oscillations of the
persistent current in the ring. Moreover, in the wide quantum rings the effects of the
penetration of magnetic field into the conducting region arise. These effects result in the
breakdown of the linear relation between the persistent current and the magnetization \cite{TI99}
and the appearance of dHvA-type oscillations of the magnetization of the ring.
\cite{GMSH,TI99,BL-95}

In recent years, the effect of the surface curvature on the spectral, magnetic, and transport
properties of nanostructures has attracted a substantial interest.
\cite{EM,Batista,Carey98,BGM-00,Leadbeater} The recent progress in nanotechnology has made it
possible to produce curved 2D layers \cite{Prinz00} and nanometer-size objects of desired shapes.
\cite{Prinz01} In particular, an original technique developed in Refs.
\onlinecite{Prinz00,Prinz01} enables fabricating nanotubes, quantum rolls, rings, and spiral-like
strips of precisely controllable shapes and dimensions.

In this paper, we study the effect of surface curvature on the magnetic moment and persistent
currents in a 2D quantum ring. Noninteracting spinless electrons in a 2D ring with a constant
negative curvature are considered. The 2D ring is placed in an orthogonal magnetic field and an
AB flux piercing through the center of the ring. The considered model of such a nanostructure is
very flexible: both the radius and the width of the ring can be adjusted independently by
suitably choosing the two parameters of the confining potential. Moreover, 1D rings and curved
quantum dots can be described at peculiar values of the parameters. For the mathematical
description of the considered system, we use a ring domain in a manifold of negative constant
curvature (in mathematical literature also known as the Lobachevsky plane); for the geometric
confinement of the ring, we choose a kind of the soft-wall potential (see the next section for
details). It is significant that the problem of the physics on the surfaces of constant curvature
has a deep relation with some interesting problems, like the quantum chaos
\cite{Gutzwiller,Stockmann}, influence of the negative curvature on  the Berry phase
\cite{Albeverio} and on the spectrum of the magneto-Bloch electron. \cite{Bruning} In recent
years, the QHE on the Lobachevsky plane is a subject of current interest.
\cite{Iengo,Ali96,Carey98,BGM-03}

\section{Electron energy spectrum}
Let us consider a 2DEG on a surface $L$ of constant negative curvature (the Lobachevsky plane)
subjected to an orthogonal (to the surface) magnetic field that is the superposition of a uniform
magnetic field $\mathbf{B}$ and the field of an Aharonov--Bohm solenoid with the flux
$\Phi_{AB}$. We employ the Poincar\'{e} realization for $L$ identifying it with the complex disc
$\{z\in\mathbb{C}:|z|<2a\}$ endowed with the metric
\[
ds^2=\frac{dr^2+r^2d\varphi^2}{[1-(r/2a)^2]^2},
\]
where $a$ is the radius of curvature; $(r,\varphi)$ are the polar coordinates in the plane
$\mathbb{C}:\ z=r\exp(i\varphi)\ (0<r<2a,\ 0\le\varphi<2\pi)$.

The vector potential $\mathbf{A}$ may be represented as the sum of
two terms: $\mathbf{A}=\mathbf{A}_1+\mathbf{A}_2$, where
\[
\mathbf{A}_1=\left(0,\frac{Br}{2[1-(r/2a)^2]}\right)
\]
is the vector potential of the uniform component and
$\mathbf{A}_2=(0,\Phi_{AB}/2\pi r)$ is the vector potential of the
AB flux.

The Hamiltonian of such a system is given by
\begin{widetext}
\begin{eqnarray}\nonumber
H_0&=&\frac{\hbar^2}{2m^*a^2}\left\{ -a^2\left[1-\left(\frac{r}{2a}\right)^2\right]^2\left[
\frac{\partial^2}{\partial r^2}+\frac1r\frac{\partial}{\partial
r}+\frac{1}{r^2}\left(\frac{\partial}{\partial\varphi}+i\frac{\Phi_{AB}}{\Phi_0}\right)^2\right]\right\}-\\
\label{eq:H0} &&
-i\frac{\hbar\omega_c}{2}\left[1-\left(\frac{r}{2a}\right)^2\right]\left(\frac{\partial}{\partial\varphi}+i\frac{\Phi_{AB}}{\Phi_0}\right)+
\frac{m^*\omega_c^2a^2}{2}\left(\frac{r}{2a}\right)^2-\frac{\hbar^2}{8m^*a^2},
\end{eqnarray}
\end{widetext}
where $m^*$ is the effective electron mass and $\omega_c$ is the cyclotron frequency. The last
term in Eq.~(\ref{eq:H0}) is the surface potential \cite{Costa} which arises from the surface
curvature. \cite{Albeverio}

We consider a 2D ring on $L$ defined by a radial confining
potential
\begin{equation}
\label{eq:V}
V(r)=\lambda_1r^2+\frac{\lambda_2}{r^2}[1-(r/2a)^2]^2-V_0,
\end{equation}
where $\lambda_1, \lambda_2$ are the parameters of the potential,
$V_0=-\lambda_2/2a^2+2\sqrt{\lambda_2[\lambda_1+\lambda_2/(2a)^4]}$.
The potential has the minimum $V(r_0)=0$ at
\[
r_0=\left(\frac{\lambda_2}{\lambda_1+\lambda_2/(2a)^4}\right)^{1/4}.
\]
Hence $r_0$ defines the average radius of the ring. It is easy to show, that for $r\simeq r_0$
the confining potential has the parabolic form
\[
V(r)\simeq\frac12 m^*\omega_0^2(r-r_0)^2,
\]
where the frequency
$\omega_0=\sqrt{8[\lambda_1+\lambda_2/(2a)^4]/m^*}$ characterizes
the strength of the transverse confinement.

The outer $r_+$ and inner $r_-$ radiuses of the ring (and,
therefore, its width $\Delta r=r_+-r_-$) can be expressed in terms
of the Fermi energy $E_F$:
\[
r_\pm=\left(\frac{V_0+E_F\pm\sqrt{2E_FV_0+E_F^2}}{2\lambda_1+\lambda_2/8a^4}\right)^{1/2}.
\]

Note that in the limit of zero curvature $(a\to\infty)$, the confinement potential (\ref{eq:V})
has the form
\[
V(r)\mathop{\longrightarrow}\limits_{a\to\infty}\lambda_1r^2+\frac{\lambda_2}{r^2}
-2\sqrt{\lambda_1\lambda_2}.
\]
As shown in Refs.~\onlinecite{BL-95,KBH,TI99,TI96}, this potential provides a good description of
the confinement for actual mesoscopic rings.

It is convenient to express the parameters of the confining potential in terms of $r_0$ and
$\omega_0$. Note that the model defined by Eq.~(\ref{eq:V}) is very flexible. It can also be used
to describe an 1D quantum ring ($r_0=\mathrm{const},\ \omega_0\to\infty$) and a 2D quantum dot
($r_0=0$).

The Schr\"{o}dinger equation for electrons on the surface of constant negative curvature with the
confining potential defined by Eq.~(\ref{eq:V}) reads
\[
(H_0+V(r))\psi(r,\varphi)=E\psi(r,\varphi)\,.
\]
By substituting the wave function $\psi(r,\varphi)=(\exp(im\varphi)/\sqrt{2\pi})f_m(r)$ and by
changing the variable $x=1/[1-(r/2a)^2]$, the Schr\"{o}dinger equation is simplified to
\[
(H_m-2m^*a^2E/\hbar^2)f_m(x)=0,
\]
where
\begin{eqnarray}\nonumber
H_m&=&-\frac{d}{dx}x(x-1)\frac{d}{dx}+\frac{M^2}{4}\frac{1}{x-1}
-\frac{{m^*}^2\omega_m^2a^4}{\hbar^2}\frac1x+\\
\label{eq:H_m}
&&+\frac{{m^*}^2a^4}{\hbar^2}\left\{\omega_c^2+\omega_0^2\left[1-\left(\frac{r_0}{2a}\right)^2\right]^2\right\}-\frac14,
\end{eqnarray}
\begin{eqnarray}
\label{eq:MLob}
M&=&\sqrt{(m+\phi_{AB})^2+(m^*\omega_0r_0^2/2\hbar)^2},\\
\label{eq:OmegaLob}
\omega_m&=&\sqrt{[\omega_c-\hbar(m+\phi_{AB})/2m^*a^2]^2+\omega_0^2},
\end{eqnarray}
$\phi_{AB}=\Phi_{AB}/\Phi_0$ is the number of Aharonov --- Bohm
flux quanta.

After some algebra we find that the spectrum of $H_m$ consists of two parts: a discrete spectrum
in the interval $(0,E_0)$ and a continuous one in the interval $[E_0,\infty)$, where
$E_0=m^*\omega_c^2a^2/2+m^*\omega_0^2a^2[1-(r_0/2a)^2]^2/2$ is the lower bound of the continuous
spectrum.

The discrete spectrum consists of finite numbers of eigenvalues
\begin{eqnarray}
\nonumber
E_{nm}&=&\hbar\omega_m\left(n+\frac12+\frac{M}{2}\right)+\frac{\hbar\omega_c}{2}(m+\phi_{AB})\\
\nonumber&&-\frac{m^*\omega_0^2r_0^2}{4}-\frac{\hbar^2}{2m^*a^2}\left[\frac12(m+\phi_{AB})^2\right.\\
\label{eq:ApSpectr}&&\left.+\left(n+\frac12\right)^2+\left(n+\frac12\right)M\right],
\end{eqnarray}
where
\begin{equation}
\label{eq:n_max} n\in\mathbb{N}:0\le
n<m^*\omega_ma^2/\hbar-M/2-1/2.
\end{equation}
The corresponding orthonormal eigenfunctions of $H_m$ are given by
\begin{eqnarray}\nonumber
f_{nm}(x)&=&C_{nm}(x-1)^{\beta_m}
x^{n-\alpha_m}\\
\label{eq:f_nm}&&\times F(-n,-n+2\alpha_m;2\alpha_m-2\beta_m-2n;1/x),
\end{eqnarray}
where $\alpha_m=\omega_m/\Omega,\ \beta_m=M/2$.

 Using the properties of hypergeometric functions, one can find that
normalization constants $C_{nm}$:
\begin{eqnarray*}
|C_{nm}|^2&=&2^{2\alpha_m-2\beta_m-2n-2}\frac{\Gamma(2\alpha_m-n)\Gamma(2\alpha_m-2\beta_m-n)}{a^2[\Gamma(2\alpha_m-2\beta_m-2n)]^2}\\
&&\times\frac{(2\alpha_m-2\beta_m-2n-1)}{\Gamma(1+2\beta_m+n)n!}\,,
\end{eqnarray*}
where $\Gamma(x)$ is the Euler Gamma-function.

The continuous spectrum of $H_m$ is defined by
\begin{equation}
\label{eq:Spectrum}
E_{\nu}=\frac{m^*\omega_c^2a^2}{2}+\frac{m^*\omega_0^2a^2}{2}[1-(r_0/2a)^2]^2+
\frac{\hbar^2}{2m^*a^2}\nu^2,\ \nu\in\mathbb{R}.
\end{equation}
The corresponding orthonormal eigenfunctions are given by
\begin{eqnarray*}
f_{\nu m}(x)&=&C_{\nu m}x^{\alpha_m}(x-1)^{\beta_m}F(\alpha_m+\beta_m+i\nu+1/2,\\
&&\alpha_m+\beta_m-i\nu+1/2;1+2\beta_m;1-x),\\
\end{eqnarray*}
where
\begin{eqnarray*}
C_{\nu m}&=&\frac{(\sinh2\pi\nu)^{1/2}}{2\pi a\Gamma(1+2\beta_m)}\left|\Gamma(-\alpha_m+ \beta_m-i\nu+1/2)\right.\\
&&\times\left.\Gamma(\alpha_m+ \beta_m-i\nu+1/2)\right|.
\end{eqnarray*}
As follows from Eq.~(\ref{eq:Spectrum}), the lower bound of the continuous spectrum is a
quadratic function of the uniform component of the magnetic field and independent of the AB flux.
There is a finite number of discrete eigenvalues of $H_m$ below this bound.

Note that the width of the quantum ring $\Delta r<r_0$ and for the quantum dot $\Delta r<2a$.
Therefore, $E_F\ll m^*\omega_0^2a^2[1-(r_0/2a)^2]^2/2$ for these nanostructures. In this case, as
can be seen from Eq.~(\ref{eq:Spectrum}), the continuous spectrum is much higher than the Fermi
energy for the quantum ring and dot. Hence, at low temperatures, there is no contribution of the
continuous spectrum of electrons to magnetic and transport properties  of these nanostructures.

As can be seen from Eq.~(\ref{eq:ApSpectr}), the effect of an AB flux on the energy spectrum and
the wave function of electrons is to shift $m\to m+\phi_{AB}$. Note that, in contrast to the case
of an 1D ring, an AB flux changes not only the phases of the wave functions but also the
trajectory of electron states in a 2D quantum ring (see Eq.~(\ref{eq:f_nm})), which results in a
non-parabolic dependence of the energy spectrum on the AB flux. \cite{TI96}

As can be seen from Eq.~(\ref{eq:ApSpectr}), the energy levels with the same quantum number $n$
form a subband. The energy spectrum of the ring is a periodic function of $\Phi_{AB}$ with the
period  $\Phi_0$. While the energy spectrum of the ring is an aperiodic function of $B$. Thus,
the effect of the penetration of a magnetic field into the conducting region of the ring is to
the appearance of the dependence of a subband minima on a magnetic field and to the asymmetry of
the subband dispersion about the subband minimum. \cite{TI96} Note that the effect of the surface
curvature is to the additional subband asymmetry. In the following we consider the case of
$\phi_{AB}=0$ only.

As can be seen from Eq.~(\ref{eq:ApSpectr}), at zero magnetic fields, the minima of all subbands
are at $m=0$. At non-zero magnetic fields, all subband minima lie at $m=m_0$, where
\begin{equation}
\label{eq:m_0} m_0=\frac{m^*\omega_cr_0^2}{2\hbar[1-(r_0/2a)^2]}.
\end{equation}
Note that  $m_0$ is the number of quantum flux circled by a ring with an effective radius $r_0$
as well as for the case of the flat surface. \cite{TI96} The dependence of an subband minimum on
a magnetic field is given by
\begin{equation}
\label{eq:E_nm0}
E_{n,m_0}=\hbar\widetilde{\omega}\left(n+\frac12\right)-\frac{\hbar^2}{2m^*a^2}
\left(n+\frac12\right)^2,
\end{equation}
where $\widetilde{\omega}=\sqrt{\omega_c^2+\omega_0^2[1-(r_0/2a)^2]^2}$. Note that in the limit
of zero curvature (the case of the flat surface) we get the following formula for an subband
minimum:
\begin{equation}
\label{eq:E_nm0_flat}
E_{n,m_0}\mathop{\longrightarrow}\limits_{a\to\infty}\hbar\omega\left(n+\frac12\right),
\end{equation}
where $\omega=\sqrt{\omega_c^2+\omega_0^2}$. Note that, in the case of a ring on the surface of
constant negative curvature, $r_0<2a$, therefore, the hybrid frequency $\widetilde{\omega}$ is
less than the one corresponding to the case of the flat surface ($\omega$). The survace curvature
decreases the contribution of the term which is due to the ring width in $\widetilde{\omega}$.
Hence, the decrease of the transverse confinement of a ring is one of manifestations of the
surface curvature. As can be seen from Eq.~(\ref{eq:E_nm0_flat}), the spacing between the bottoms
of neighboring subbands is $\hbar\omega$.  From Eq.~(\ref{eq:E_nm0}) it can be easily shown that
for the case of the Lobachevsky plane this spacing is less than that for the flat surface and
depends on the subband index. It can be seen from Eq.~(\ref{eq:E_nm0}) that the bottoms of
subbands are increasing with a magnetic field. Moreover, the magnetic field dependence of the
subband bottoms is stronger for the higher subbands. The decrease of the ring width shifts the
subbands to higher energies, increases the subband spacings, and weakens the dependence of the
subband bottom on a magnetic field. As can be seen from Eqs.~(\ref{eq:E_nm0}) and
(\ref{eq:E_nm0_flat}), the surface curvature leads to the converse effects: the decrease of the
curvature radius shifts the subbands to lower energies, decreases the subband spacings, and
strengthens the dependence of the subband bottoms on a magnetic field.

Let us study the limiting cases. Firstly, we consider the case of an 1D ring
($r_0=\mathrm{const},\ \omega_0\to\infty$). The energy spectrum of the nanostructure at
$\omega_0\to\infty$ can be found in the same way as in Ref.~\onlinecite{TI96}:
\begin{eqnarray}
\nonumber E_{n,m}&=&\hbar\widetilde{\omega}\left(n+\frac12\right)-\frac{\hbar^2}{2m^*a^2}
\left(n+\frac12\right)^2\\
\label{eq:ring}&&+\frac{\hbar^2}{2m^*r_0^2}\left[1-\left(\frac{r_0}{2a}\right)^2\right]^2
(m-m_0)^2,
\end{eqnarray}
where $m_0$ is defined by Eq.~(\ref{eq:m_0}). As can be seen from Eq.~(\ref{eq:ring}), electrons
occupy the lowest subband only. The subband bottom is independent of a magnetic field and the
subband energy spectrum is symmetric about $m_0$. Comparing Eq.~(\ref{eq:ring}) and that for the
case of the flat surface, we find that the surface curvature shifts the subband minima to the
lower magnetic field and weakens the magnetic field dependence of the electron energy.

Secondly, we consider the case of a quantum dot ($r_0=0$). Assuming that $r_0=0$ in
Eq.~(\ref{eq:ApSpectr}), we find the energy spectrum of electrons in a quantum dot on the surface
of constant negative curvature
\begin{eqnarray}
\nonumber E_{n,m}&=&\hbar\omega_m\left(n+\frac12+\frac{|m|}{2}\right)+\frac{\hbar\omega_cm}{2}-
\frac{\hbar^2}{2m^*a^2}\\
\label{eq:dot}&&\times\left[\frac{m^2}{2}+\left(n+\frac12\right)^2+\left(n+\frac12\right)|m|\right].
\end{eqnarray}
It is well known \cite{TI99} that the energy spectrum of an isotropic 2D quantum dot on the flat
surface is degenerate at zero magnetic fields (the $n$-th level is $n$-fold degenerate). The low
magnetic field lifts the degeneracy of the levels, whereas, at high magnetic fields, the energy
levels form the Landau subbands. As can be seen from Eq.~(\ref{eq:dot}), the surface curvature
shifts the energy levels to the lower energy. Moreover, the shift is more essential for the
levels with higher $n$ or $m$. Furthermore, the surface curvature reduces the level degeneracy at
zero magnetic fields. The level with $m=0$ is nondegenerate, whereas the levels with $m\ne 0$ are
two-fold degenerate. As can be seen from Eq.~(\ref{eq:dot}), the hybrid frequency $\omega_m$
depends on the quantum number $m$, therefore, for the levels with small $|m|$, the curvature has
negligible effect on the behavior of the levels in magnetic fields. On the contrary, for the
levels with large $|m|$, the curvature essentially changes the hybrid frequency and changes the
behavior of the levels in a magnetic field.

\section{Magnetic moment}
The magnetic moment of a system containing a fixed number of electrons is given by \cite{BGM-00}
\begin{equation}
\label{eq:MN} \mathcal{M}=-\left(\frac{\partial F}{\partial B}\right)_{T,N}=
\sum_{n,m}\mathcal{M}_{n,m}f_0(E_{n,m}),
\end{equation}
where $F$ is the free energy, $N$ is the number of electrons,
\begin{equation}
\label{eq:Mnm}
\mathcal{M}_{n,m}=-\frac{\partial E_{n,m}}{\partial B}
\end{equation}
is the magnetic moment of the $(n,m)$th state. The chemical potential of a system is determined
completely by the normalization condition
\[
N=\sum_{n,m}f_0(E_{n,m}).
\]
For the case of a quantum ring on the Lobachevsky plane, substituting Eq.~(\ref{eq:ApSpectr})
into Eq.~(\ref{eq:Mnm}), we get
\begin{eqnarray}
\nonumber
\frac{\mathcal{M}_{n,m}}{\mu_B}&=&-\frac{m_e}{m^*}\Bigg[m+\phi_{AB}+\frac{\omega_c-\hbar
(m+\phi_{AB})/2m^*a^2}{\omega_m}\\
\label{eq:Mring}&&\times(2n+1+M)\Bigg],
\end{eqnarray}
where $M$ and $\omega_m$ are defined by Eq.~(\ref{eq:MLob}) and Eq.~(\ref{eq:OmegaLob})
respectively.

\begin{figure}
\includegraphics[width=8.6cm]{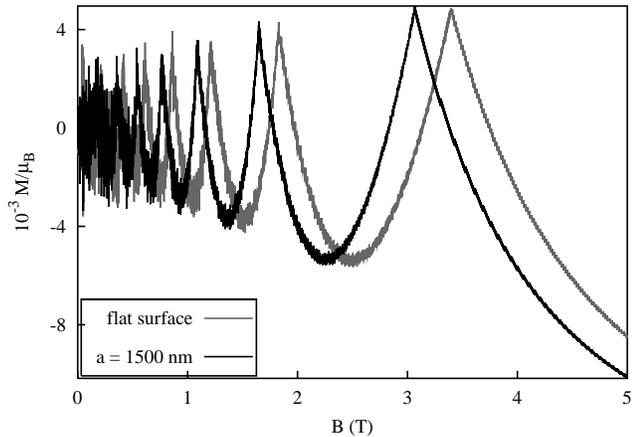}
\caption{\label{fig:5_5_7}The magnetic moment of a 2D quantum ring as a function of a magnetic
field; $N=1000$, $r_0=800\;$nm, $\omega_0 = 1.5\times 10^{12}\;$s$^{-1}$, $\phi_{AB}=0$, and
$T=0\;$ K.}
\end{figure}

\begin{figure}
\includegraphics[width=8.6cm]{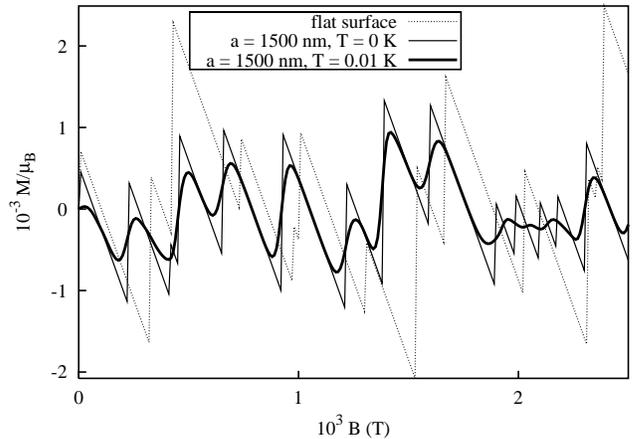}
\caption{\label{fig:5_5_8}The magnetic moment of a 2D quantum ring as a function of a magnetic
field (the case of low magnetic fields); $N=1000$, $r_0=800\;$nm, $\omega_0 = 1.5\times
10^{12}\;$s$^{-1}$, and $\phi_{AB}=0$.}
\end{figure}

\begin{figure}
\includegraphics[width=8.6cm]{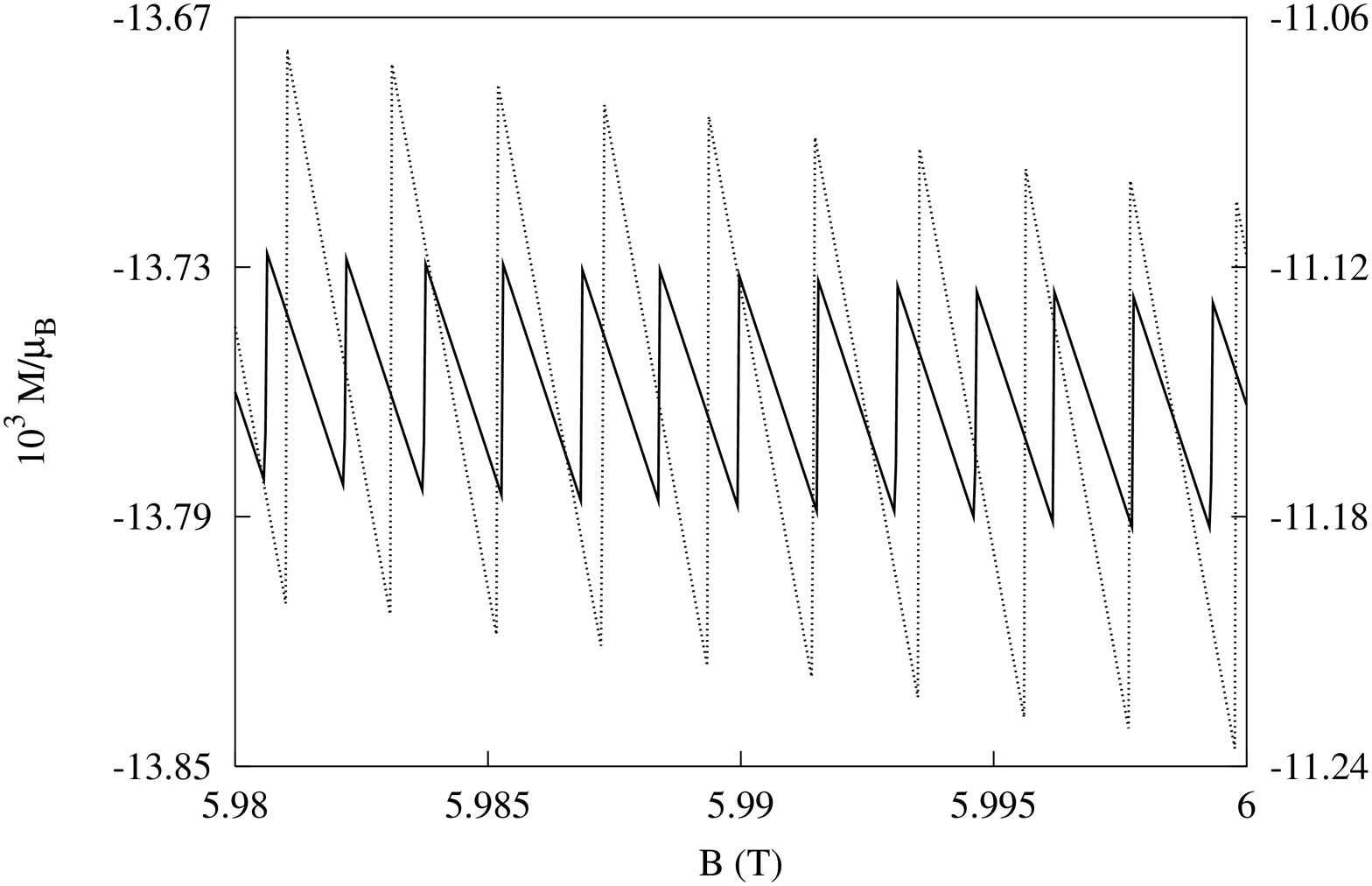}
\caption{\label{fig:5_5_9}The dependence of the magnetic moment of a 2D quantum ring on a
magnetic field (the case of high magnetic fields). The magnetic moment of a 2D ring on the
Lobachevsky plane with the radius of curvature $a=800\;$nm is plotted with a solid line (left
ordinate). The magnetic moment of a 2D ring on the flat surface is plotted with a dotted line
(right ordinate); $N=1000$, $r_0=800\;$nm, $\omega_0 = 1.5\times 10^{12}\;$s$^{-1}$,
$\phi_{AB}=0$, and $T=0\;$K.}
\end{figure}

As shown in Fig.~\ref{fig:5_5_7}, the magnetic moment of a 2D quantum ring on the Lobachevsky
plane as well as on the flat surface \cite{BL-95,TI99} has a complex oscillation pattern:
oscillations of the AB-type are superimposed on oscillations of the dHvA-type. The amplitude of
AB-type oscillations is strongly suppressed by increasing magnetic-field strength, whereas the
amplitude of dHvA-type oscillations is increased with a magnetic field. At low magnetic fields,
the amplitudes of these types of oscillations are of the same order of magnitude, and the
superimposition of these oscillations leads to the appearance of a beating pattern in
$\mathcal{M}(B)$ (Fig.~\ref{fig:5_5_8}). At high magnetic fields, the amplitude of AB-type
oscillations is much smaller than that of dHvA-type oscillations (Fig.~\ref{fig:5_5_7}).
Moreover, the AB-type oscillations are almost periodic in the strong-magnetic-field regime (Fig.
~\ref{fig:5_5_9}).

Note that the AB-type oscillations of the magnetic moment arise from the electron level
crossings. Whereas, the dHvA-type oscillations arise from singularities in the electron density
of states. It can be shown that the electron density of states is larger at the subband bottoms.
Therefore, the maximums of dHvA-type oscillations arise when the chemical potential crosses the
subband bottoms.

Let us consider the effect of the surface curvature on the magnetic moment of a ring. As
mentioned above, the decrease of subband spacings is the one of manifestations of the surface
curvature. This leads to the decrease of the spacing between neighboring maxima of dHvA-type
oscillations (Fig.~\ref{fig:5_5_7}). Moreover, in the limit of low magnetic fields, the number of
subbands below the Fermi energy is larger than that for the case of the flat surface. Therefore,
the number of oscillations increases and the maximum of the amplitude of oscillations decreases
with the increasing of surface curvature (Fig.~\ref{fig:5_5_8}).

The dependence of the hybrid frequency on the magnetic quantum number is the another
manifestation of the surface curvature (Eq.~(\ref{eq:OmegaLob})). As follows from this equation,
the surface curvature weakens the dependence of the energy levels on a magnetic field and
decreases the level spacing. The former decreases the amplitude of AB-type oscillations, while
the latter decreases the period of these oscillations (Fig.~\ref{fig:5_5_9}).
As shown in Fig.~\ref{fig:5_5_9}, the monotonic part of the magnetic moment for a ring on the
Lobachevsky plane is below that for the flat surface.

Note that temperature results in smearing of the oscillation maxima and decreasing the
oscillation amplitude (Fig.~\ref{fig:5_5_8}).

Let us consider particular cases. In the limit of an 1D ring, substituting Eq.~(\ref{eq:ring})
into Eq.~(\ref{eq:Mnm}), we get for the magnetic moment
\begin{equation}
\label{eq:1Dring}
\frac{\mathcal{M}_{n,m}}{\mu_B}=\frac{m_e}{m^*}\left[1-\left(\frac{r_0}{2a}\right)^2\right]
\left(m+\frac{\Phi_{AB}}{\Phi_0}-\frac{\Phi}{\Phi_0}\right),
\end{equation}
where $\Phi=BS$ is the magnetic flux penetrating through a ring, $S=\pi r_0^2/[1-(r_0/2a)^2]$ is
the surface area circled by a ring with an effective radius $r_0$. As can be seen from this
equation, the magnetic moment is a periodic function of a magnetic field with a period $\Phi_0$.
As follows from the analysis of the energy spectrum of electrons in an 1D ring, the surface
curvature decreases the amplitude of the magnetic moment oscillations.

In the limit of a quantum dot, assuming that $r_0=0$ in Eq.~(\ref{eq:Mring}), we obtain
\begin{eqnarray}
\nonumber
\frac{\mathcal{M}_{n,m}}{\mu_B}&=&-\frac{m_e}{m^*}\Bigg[m+\phi_{AB}+\frac{\omega_c-\hbar
(m+\phi_{AB})/2m^*a^2}{\omega_m}\\
\label{eq:Mdot}&&\times(2n+1+|m+\phi_{AB}|)\Bigg].
\end{eqnarray}

\begin{figure}
\includegraphics[width=8.6cm]{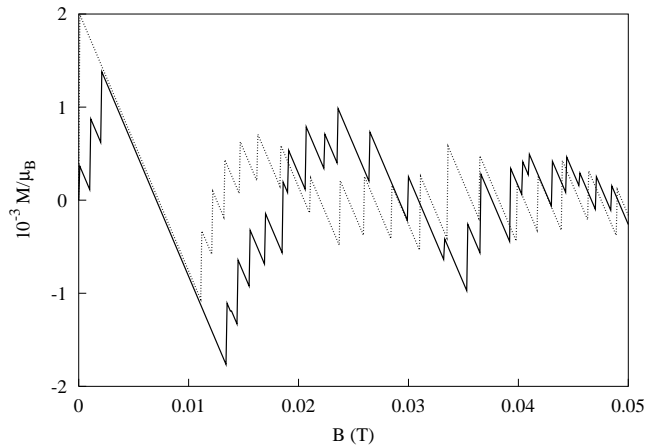}
\caption{\label{fig:5_5_11}The dependence of the magnetic moment of a quantum dot on a magnetic
field (the case of low magnetic fields). The magnetic moment of a dot on the Lobachevsky plane
with the radius of curvature $a=800\;$nm is plotted with a solid line. The magnetic moment of a
dot on the flat surface is plotted with a dotted line; $N=500$,
$\omega_0=1.5\times10^{12}\;$s$^{-1}$, $\phi_{AB}=0$, and $T=0\;$Š.}
\end{figure}

\begin{figure}
\includegraphics[width=8.6cm]{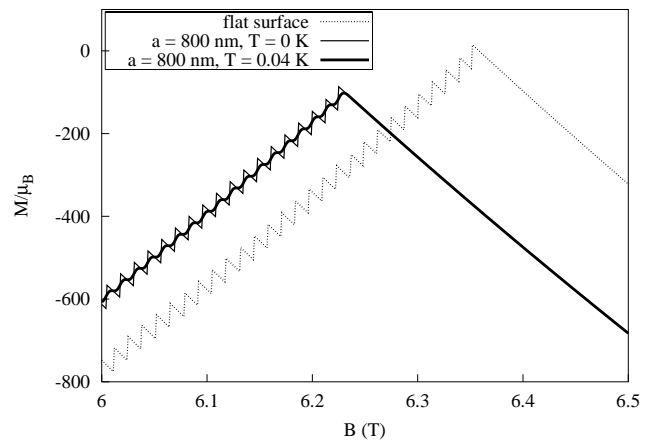}
\caption{\label{fig:5_5_12}The magnetic moment of a quantum dot as a function of a magnetic field
(the case of a high magnetic fields); $N=500$, $\omega_0=1.5\times10^{12}\;$s$^{-1}$, and
$\phi_{AB}=0$.}
\end{figure}

As mentioned above, for the zero magnetic fields, the energy levels of an isotropic 2D quantum
dot on the flat surface are highly degenerate. The low magnetic field lifts the degeneracy of the
levels, and near the Fermi energy more electrons will occupy the states with $m<0$, which have a
lower energy than the $m\ge0$ states. Therefore, the magnetic moment of the quantum dot on the
flat surface has a large value at a weak magnetic field. \cite{TI99} Further increase of a
magnetic field leads to complex oscillations of $\mathcal{M}(B)$, which arise from the
superimposition of the AB-type oscillations on the dHvA-type oscillations.
\begin{figure}
\includegraphics[width=8.6cm]{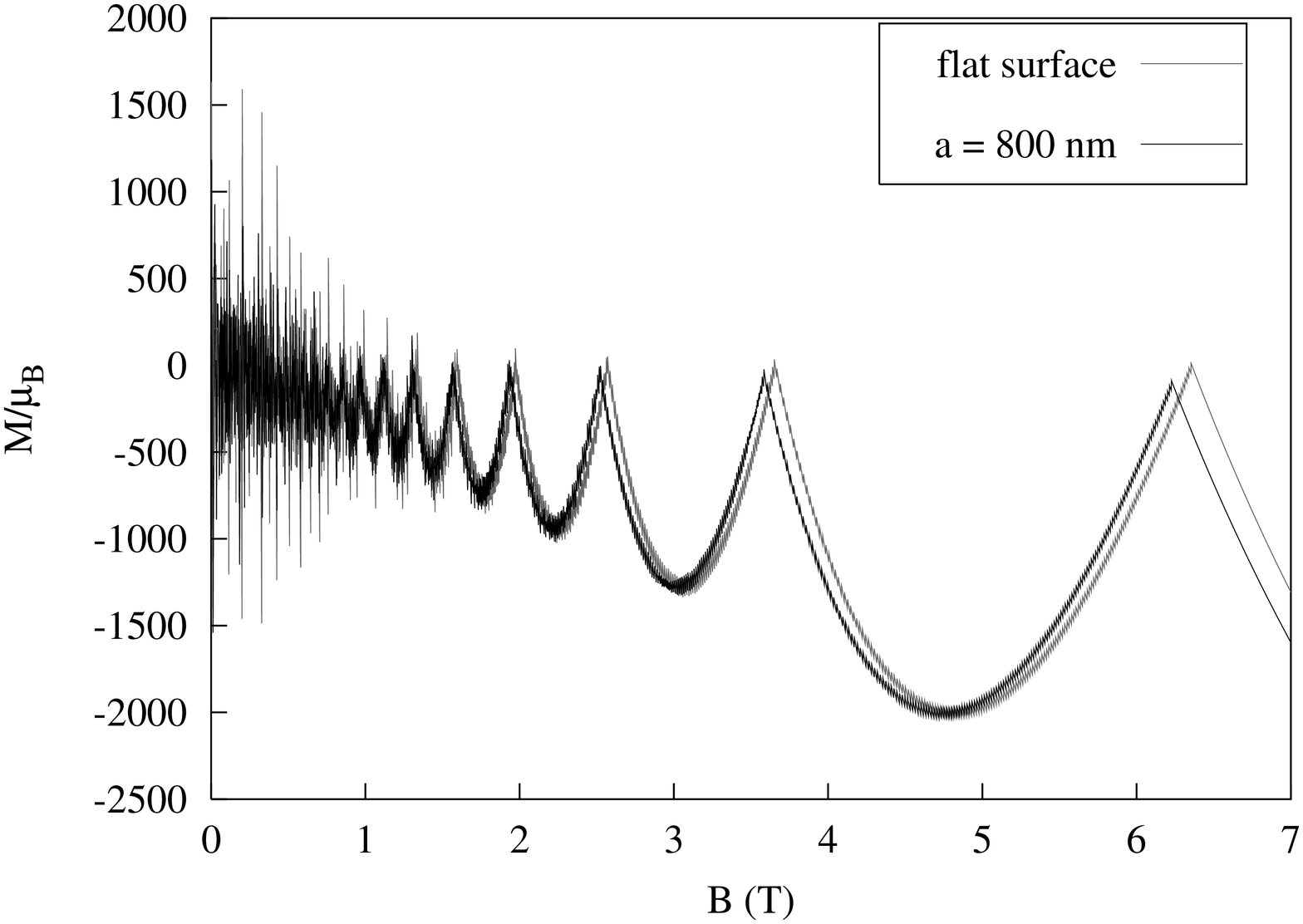}
\caption{\label{fig:5_5_13}The magnetic moment of a quantum dot as a function of a magnetic
field; $N=500$, $\omega_0=1.5\times10^{12}\;$s$^{-1}$, $\phi_{AB}=0$, and $T=0\;$K.}
\end{figure}
In Fig.~\ref{fig:5_5_13} we show the dependence $\mathcal{M}(B)$ for the Lobachevsky plane as
well as for the flat surface. As follows from the behavior of the electron levels in a magnetic
field, the surface curvature decreases the period and the amplitude of dHvA-type oscillations
(Fig.~\ref{fig:5_5_13}). Since the surface curvature reduces the level degeneracy at zero
magnetic fields, the magnetic moment of a quantum dot on the Lobachevsky plane is lower than that
for the flat surface at weak magnetic fields. The new jumps in the field-dependence of the
magnetic moment arise with increasing the magnetic field, which are due to the crossings of
degenerated levels at $B=0$ in the case of the flat surface (Fig.~\ref{fig:5_5_11}).
At high magnetic fields, when the lowest subband is occupied only, there is no level crossings,
therefore, AB-type oscillations vanish (Fig.~\ref{fig:5_5_12}).

\section{Persistent currents}

Let us consider the persistent current of a 2D quantum ring on the Lobachevsky plane. We study
the relation between the persistent current and the magnetic moment of a ring and we examine the
effect of the surface curvature on the persistent current of a ring. When wave functions of a
system are zero at $r=0$, the persistent current can be calculated using the following equation
\cite{BayersYang,Bloch}
\begin{equation}
\label{eq:Current}
I=-c\left(\frac{\partial F}{\partial \Phi_{AB}}\right)_{T,N}=\sum_{n,m}I_{n,m}f_0(E_{n,m}),
\end{equation}
where
\begin{equation}
\label{eq:Inm}
I_{n,m}=-c\frac{\partial E_{n,m}}{\partial \Phi_{AB}}.
\end{equation}

\begin{figure}
\includegraphics[width=8.6cm]{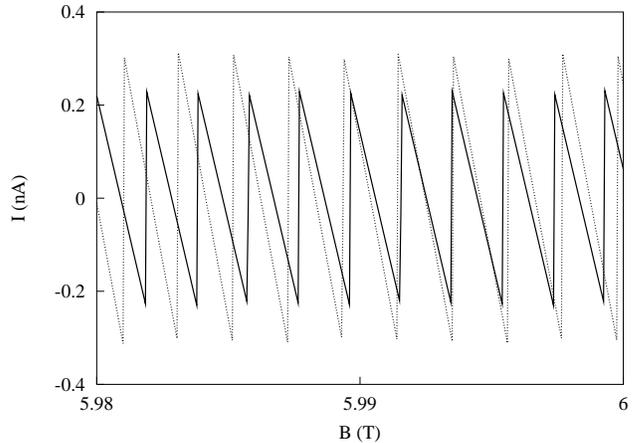}
\caption{\label{fig:5_5_15}The dependence of the persistent currents of a 2D ring on a magnetic
field (the case of high magnetic fields). The persistent currents in a ring on the Lobachevsky
Plane with the radius of curvature $a=1500\;$nm is plotted with a solid line. The persistent
currents in a ring on the flat surface is plotted with a dotted line; $N=1000$, $r_0=800\;$nm,
$\omega_0=1.5\times10^{12}\;$s$^{-1}$, $\phi_{AB}=0$, and $T=0\;$K.}
\end{figure}

Taking into account Eq.~(\ref{eq:Mring}), the persistent current of the $(n,m)$th state is given
by
\begin{equation}\label{eq:IRing}
I_{n,m}=\frac{c}{\pi r_m^2}\left\{\mathcal{M}_{n,m}\left[1-\left(\frac{r_m}{2a}\right)^2\right]+
\frac{m_e}{m^*}\mu_B\frac{\omega_c}{\omega_m}(2n+1)\right\},
\end{equation}
where $r_m=\sqrt{2\hbar M/m^*\omega_m}$ is the effective radius of the state with quantum number
$m$. \cite{TI96,TI99} The first term in Eq.~(\ref{eq:IRing}) is the classical current in a ring
with a radius $r_m$ in a magnetic field. The second term caused by the penetration of a magnetic
field into the conducting region of a ring breaks the proportionality between the magnetic moment
and the persistent current. It can be seen from Eq.~(\ref{eq:IRing}) that only in the
weak-magnetic field limit ($\omega_c\ll\omega_0$) the magnetic moment of an electron state is
proportional to its current.
\begin{figure}
\includegraphics[width=8.6cm]{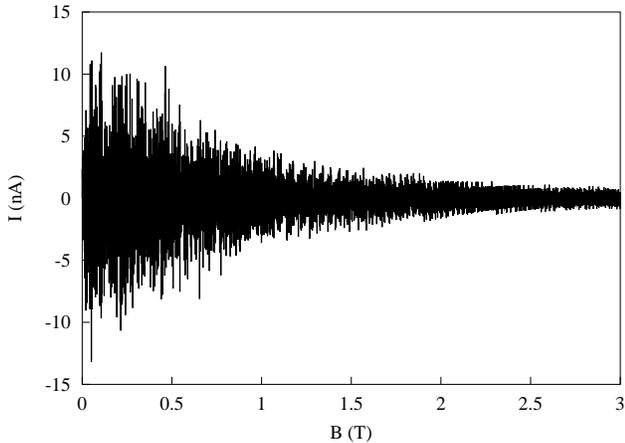}
\caption{\label{fig:5_5_16}The persistent currents of a 2D quantum ring on the Lobachevsky plane
as a function of a magnetic field; $N=1000$, $a=1500\;$nm, $r_0=800\;$nm,
$\omega_0=1.5\times10^{12}\;$s$^{-1}$, $\phi_{AB}=0$, and $T=0\;$K.}
\end{figure}
In Fig.~\ref{fig:5_5_16}, we plot the persistent current as a function of a magnetic field. As
can be seen from this figure, the persistent current shows rapid oscillations within the whole
magnetic field range. The amplitudes of the persistent current are strongly suppressed by
increasing magnetic-field strength. This is due to the fact that the oscillation amplitude
$\sim\sqrt{P}$, where $P$ is the number of the occupied subbands. \cite{TI99} This number
decreases with a magnetic field and this leads to the decrease of the oscillation amplitude. In
the limit of low magnetic fields, there are some occupied subbands. The crossings of the highest
occupied electron level with the levels of these subbands lead to the appearance of a beating
pattern in $I(B)$. As mentioned above, in this magnetic-field regime, the persistent current is
proportional to the magnetic moment. Therefore, the behavior of the persistent current is similar
to the magnetic moment behavior (Fig.~\ref{fig:5_5_8}). In the strong-magnetic-field regime, only
the lowest subband is occupied and only the crossings of levels of this subband with the highest
occupied level lead to oscillations of the persistent current. Since levels of the lowest
occupied subband are nearly equidistant, the oscillations of the persistent current are nearly
periodic in this magnetic-field regime (Fig.~\ref{fig:5_5_15}).
The period of oscillations of the persistent current is the same as for the magnetic moment
(Fig.~\ref{fig:5_5_9}). As mentioned above, the surface curvature decreases the period and the
amplitude of oscillations of the magnetic moment. As can be seen from Eq.~(\ref{eq:IRing}), the
surface curvature leads to the additional decrease of the amplitude of oscillations of the
persistent current with respect to the magnetic moment.

Note that Eq.~(\ref{eq:Inm}) also valid for electron states in a quantum dot, except for the
states with $m+\phi_{AB}=0$. This is because, when $r_0=0$, the wave function of a state
$m+\phi_{AB}=0$ has a nonzero value at $r=0$, and Eq.~(\ref{eq:Current}) no longer applies.
However, since Eq.~(\ref{eq:Inm}) applies for all states if $r_0\neq 0$, the persistent current
of the state with $m+\phi_{AB}=0$ can be obtained by taking the limit \cite{TI99}
\begin{eqnarray*}
I_{n,-\phi_{AB}}&=&\lim_{r_0\to0}\left[\lim_{m\to-\phi_{AB}}I_{n,m}\right]\\
&=&-\frac{|e|\omega_c}{4\pi}+ \frac{|e|\hbar}{4\pi
m^*a^2}\frac{\omega_c}{\omega}\left(n+\frac12\right),
\end{eqnarray*}
where $\omega=\sqrt{\omega_c^2+\omega_0^2}$.

\begin{figure}
\includegraphics[width=8.6cm]{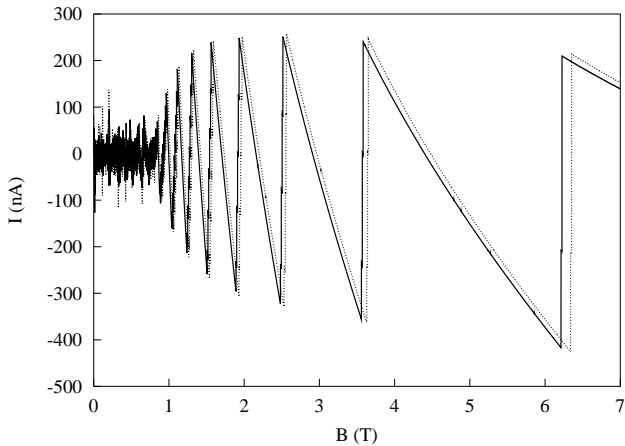}
\caption{\label{fig:5_5_17}The dependence of the persistent currents in a quantum dot on a
magnetic field. The persistent currents of a dot on the Lobachevsky plane with the radius
curvature $a=800\;$nm is plotted with a solid line. The persistent currents of a dot on the flat
surface is plotted with a dotted line; $N=500$, $\omega_0=1.5\times10^{12}\;$s$^{-1}$,
$\phi_{AB}=0$, and $T=0\;$K.}
\end{figure}

\begin{figure}
\includegraphics[width=8.6cm]{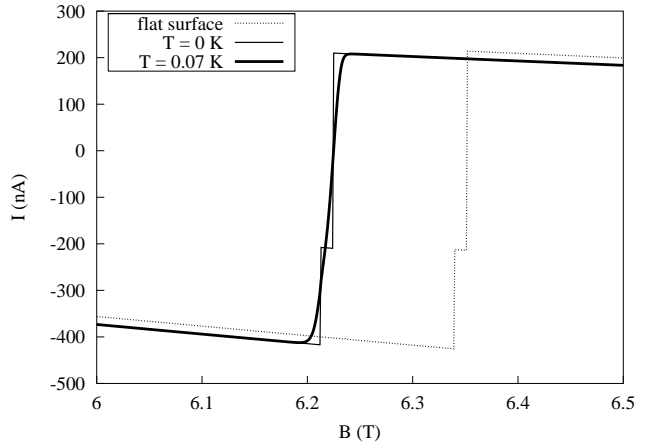}
\caption{\label{fig:5_5_19}The persistent currents of a quantum dot as a function of a magnetic
field (the case of high magnetic fields); $N=500$, $a=800\;$nm,
$\omega_0=1.5\times10^{12}\;$s$^{-1}$, and $\phi_{AB}=0$.}
\end{figure}

Note that the penetration of a magnetic field into the conducting region plays an essential role
with increasing the magnetic field. This leads to the break of the proportionality between the
magnetic moment and the persistent current of a quantum dot. The persistent current as a function
of a magnetic field exhibits rapid jumps which appear when the Fermi energy crosses the subband
bottoms (Fig.~\ref{fig:5_5_17}).
The oscillations due to the level crossings with the highest occupied level are superimposed on
these jumps. The amplitude of these oscillations decreases with a magnetic field. At high
magnetic fields, this amplitude tends to zero (Fig.~\ref{fig:5_5_19}).
As can be seen from Figs.~\ref{fig:5_5_17} and \ref{fig:5_5_19}, the surface curvature decreases
the amplitude of dHvA-type oscillations and shifts these oscillations in the low-field region.

\section{Conclusions}
The effect of the surface curvature on the magnetic moment and persistent currents of 2D quantum
rings and dots is investigated. It is shown that the surface  curvature decreases the subband
spacings. This leads to decreasing of the spacing between neighboring maxima of de Haas --- van
Alphen type oscillations (Fig.~\ref{fig:5_5_7}). Moreover, the dependence of the hybrid frequency
on the magnetic quantum number is the another manifestation of the surface curvature
(Eq.~(\ref{eq:OmegaLob})). As follows from this equation, the surface curvature weakens the
dependence of the energy levels on a magnetic field and decreases the level spacing. The former
decreases the amplitude of Aharonov --- Bohm type oscillations, while the latter decreases the
period of these oscillations (Fig.~\ref{fig:5_5_9}). Two limiting cases are considered: the case
of an 1D quantum ring and the case of a quantum dot. It is shown that the surface curvature
reduces the level degeneracy of a quantum dot at zero magnetic fields. Therefore, the magnetic
moment of a quantum dot on the Lobachevsky plane is lower than that on the flat surface at weak
fields. The persistent currents of a quantum ring and a quantum dot are studied. The relation
between the persistent current and the magnetic moment of these nanostructures is investigated.
It is shown that the surface curvature leads to the additional decrease of the amplitude of
oscillations of the persistent current with respect to the magnetic moment.

\begin{acknowledgments}
This work was supported by the INTAS Grant No.~00--257, the DFG-RAS Grant No.~436 RUS
113/572/0--2, the Russian Ministry of Education Grants Nos.~E02--3.4--370 and E02--2.0--15, the
RFBR Grants Nos.~02--01--00804 and 03--02--06006-mas.
\end{acknowledgments}

\end{document}